\documentclass[sigconf]{acmart}
\AtBeginDocument{%
  }

\copyrightyear{2026}
\acmYear{2026}
\setcopyright{cc}
\setcctype{by}
\acmConference[KDD '26]{Proceedings of the 32nd ACM SIGKDD Conference on Knowledge Discovery and Data Mining V.2}{August 09--13, 2026}{Jeju Island, Republic of Korea}
\acmBooktitle{Proceedings of the 32nd ACM SIGKDD Conference on Knowledge Discovery and Data Mining V.2 (KDD '26), August 09--13, 2026, Jeju Island, Republic of Korea}
\acmDOI{10.1145/3770855.3817750}
\acmISBN{979-8-4007-2259-2/2026/08}

\acmDOI{10.1145/3770855.3817750}
\usepackage{adjustbox}
\usepackage{amsmath}
\usepackage{amsfonts}
\DeclareMathOperator*{\topk}{topk}
\DeclareMathOperator*{\argmax}{arg\,max}

\usepackage{multirow}
\usepackage{enumitem}
\usepackage{booktabs}
\usepackage{graphicx}
\usepackage{balance}

\begin{document}

\title{Generalizing Graph Foundation Models via Hyperbolic Retrieval-Augmented Generation}

\author{Yifan Jin}
\email{jingyifan20@mails.ucas.ac.cn}
\affiliation{%
  \institution{Institute of Software, Chinese Academy of Sciences}
  \city{Beijing}
  \country{China}
}
\affiliation{
    \institution{University of Chinese Academy of Sciences}
    \city{Beijing}
    \country{China}
}

\author{Qirui Ji}
\email{jiqirui2022@iscas.ac.cn}
\affiliation{%
  \institution{Institute of Software, Chinese Academy of Sciences}
  \city{Beijing}
  \country{China}
}
\affiliation{
    \institution{University of Chinese Academy of Sciences}
    \city{Beijing}
    \country{China}
}

\author{Bin Qin}
\email{qinbin21@mails.ucas.ac.cn}
\affiliation{%
  \institution{Institute of Software, Chinese Academy of Sciences}
  \city{Beijing}
  \country{China}
}
\affiliation{
    \institution{University of Chinese Academy of Sciences}
    \city{Beijing}
    \country{China}
}

\author{Jiangmeng Li}
\authornote{Corresponding Author}
\email{jiangmeng2019@iscas.ac.cn}
\affiliation{%
  \institution{Institute of Software, Chinese Academy of Sciences}
  \city{Beijing}
  \country{China}
}

\author{Lixiang Liu}
\email{lixiang@iscas.ac.cn}
\affiliation{%
  \institution{Institute of Software, Chinese Academy of Sciences}
  \city{Beijing}
  \country{China}
}
\affiliation{
    \institution{University of Chinese Academy of Sciences}
    \city{Beijing}
    \country{China}
}

\author{Fuchun Sun}
\email{fcsun@mail.tsinghua.edu.cn}
\affiliation{%
  \institution{Tsinghua University}
  \city{Beijing}
  \country{China}}

\author{Changwen Zheng}
\email{changwen@iscas.ac.cn}
\affiliation{%
  \institution{Institute of Software, Chinese Academy of Sciences}
  \city{Beijing}
  \country{China}
}
\affiliation{
    \institution{University of Chinese Academy of Sciences}
    \city{Beijing}
    \country{China}
}

\renewcommand{\shortauthors}{Yifan Jin et al.}

\begin{abstract}
  Graph foundation models (GFMs) emerged as a dominant paradigm in graph representation learning by leveraging large-scale pre-training for cross-domain inference. However, the parameterized knowledge encoded within these models is insufficient to cope with distribution shifts, limiting their generalization ability. To mitigate this issue, retrieval-augmented generation (RAG) has been introduced to incorporate external knowledge at inference time. Nevertheless, existing RAG frameworks operating in Euclidean space suffer from a fundamental geometric limitation: the polynomial volume growth of Euclidean space is inherently mismatched with the tree-structured external knowledge bases. This mismatch leads to the loss of semantic granularity in retrieval and gives rise to the hubness phenomenon.
To address this limitation, we propose a \underline{Hy}perbolic \underline{R}etrieval-\underline{A}ugmented \underline{G}eneration (HyRAG) framework designed to enhance the generalization capabilities of GFMs. Specifically, the introduced Hyperbolic Knowledge Indexing module retains the tree-like hierarchies of the external knowledge base by modeling them within hyperbolic space. The Multi-granularity Retrieval module then provides GFMs with the global semantic anchors and local semantic nuances through coarse-grained and fine-grained knowledge retrieval, respectively. Finally, the Dual-path Fusion module achieves effective knowledge integration for graph tasks at both the feature and structural levels.
Experiments on multiple graph benchmarks demonstrate significant improvements in the zero-shot setting, highlighting the generalization of our method for robust GFMs inference.
\end{abstract}





\maketitle
\newcommand\kddavailabilityurl{https://doi.org/10.5281/zenodo.20501234}
\ifdefempty{\kddavailabilityurl}{}{
\begingroup\small\noindent\raggedright\textbf{Resource Availability:}\\
The source code of this paper has been made publicly available at \url{\kddavailabilityurl}.
\endgroup
}
\section{Introduction} \label{intro}

Graph foundation models (GFMs) have emerged as a dominant paradigm for graph representation learning, enabling unified pre-training and transfer across diverse graph tasks such as node classification \cite{graphgpt,unigraph,ofa,graphclip}, link prediction\cite{graphgpt,llaga}, and graph-level reasoning \cite{ofa,unigraph}. By leveraging pre-training on large-scale datasets, GFMs demonstrate impressive generalization capabilities. Despite these advances, the knowledge internalized by GFMs is inherently bounded by the distribution of the pre-training data\cite{Samgpt,zhao2024all}. When distributional shifts arise at inference time, the parameterized knowledge may fail to generalize, leading to degraded and unreliable inference performance\cite{li2026out}.


To mitigate this limitation, recent studies have explored retrieval-augmented generation (RAG) paradigms that allow GFMs to dynamically access external knowledge at inference time without retraining\cite{ragraph,rag4gfm}. 
For instance, RAGRAPH~\cite{ragraph} performs retrieval over a toy graph vector library and integrates the retrieved instances through a message-passing prompting mechanism to enrich the learning context. RAG4GFM~\cite{rag4gfm} introduces a multi-level indexing architecture together with a task-aware retriever to extract structurally and semantically relevant subgraphs. RAG-GFM \cite{rag-gfm} incorporates retrieval during pre-training to align semantic and structural priors.
However, we argue that RAG frameworks operating in Euclidean space face an inherent geometric limitation: 
\textit{the polynomial volume growth of Euclidean space is inherently mismatched with the tree-structured external knowledge bases,} leading to the following two issues:

\begin{figure*}
  \centering
  \includegraphics[width=2\columnwidth]{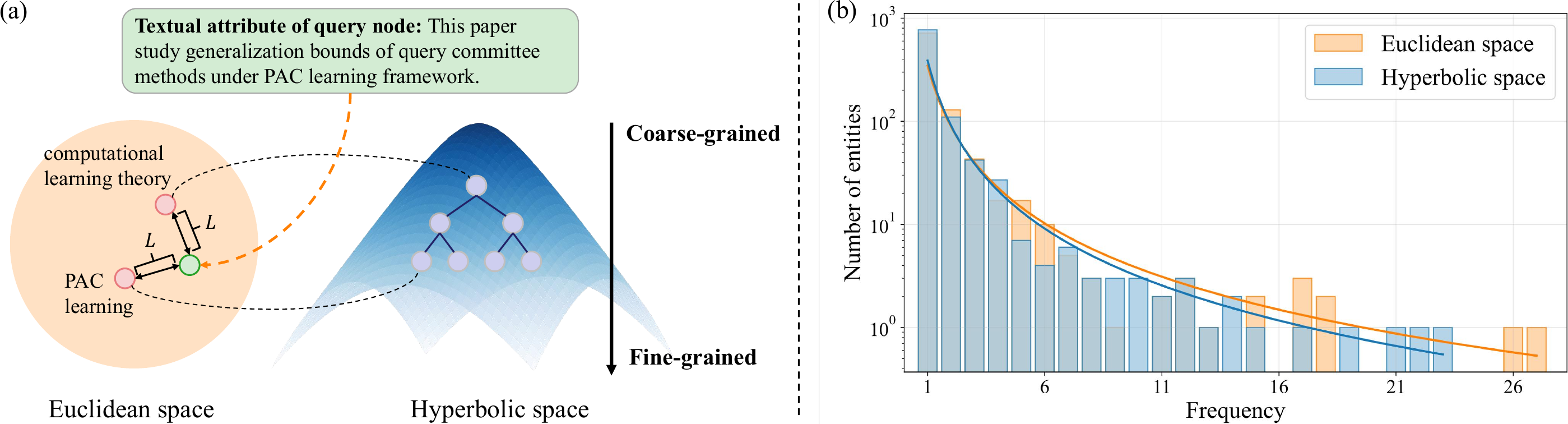}
\caption{Comparison of Euclidean space and hyperbolic space.(a) In Euclidean space (left), the query node exhibits similar distances to both coarse-grained and fine-grained concepts. Hyperbolic space (right) naturally embeds the tree-like hierarchies. (b) Node frequency distribution in top-$k$ retrieval results on the Cora dataset across Euclidean space and hyperbolic space, the x-axis shows the co-occurrence frequency of the entity, and the y-axis shows the number of entities (log scale).
}
  \label{fig:motivation}
  \vspace{-0.3cm}
\end{figure*}
 1) \textbf{Loss of semantic granularity.} 
External knowledge bases are commonly organized as tree-structured hierarchies with exponential growth, where root nodes represent coarse-grained concepts and leaf nodes correspond to fine-grained concepts. These two levels of granularity provide complementary information for graph tasks.
Specifically, for the classification of the query node in Figure~\ref{fig:motivation}(a), the coarse-grained concept ``computational learning theory'' situates the semantic context within the theoretical domain, serving as a global semantic anchor. In contrast, the fine-grained concept ``PAC learning'' offers local semantic nuances by specifying the concrete learning paradigm, thereby distinguishing the query from other theoretical branches.
However, Euclidean representation spaces exhibit only polynomial volume growth with respect to the radius, which fundamentally mismatches the exponential expansion of hierarchical knowledge and inevitably compresses concepts from different hierarchical depths into proximal regions of the embedding space.
As a consequence, Euclidean-based RAG struggles to distinguish between coarse-grained and fine-grained concepts, collapsing rich hierarchical semantics into undifferentiated similarity signals. 
 As illustrated in Figure~\ref{fig:motivation}(a), the query node may appear at similar Euclidean distances to both coarse-grained and fine-grained concepts, causing them to be treated as equally informative during retrieval and fusion. Such granularity-agnostic aggregation blurs semantic roles and weakens the reliability of downstream reasoning.
 In contrast, the intrinsic geometry of hyperbolic space naturally accommodates tree-structured hierarchies, where coarse-grained concepts are embedded near the origin and fine-grained concepts reside closer to the boundary, effectively preserving granularity-sensitive semantic distinctions.

 2) \textbf{Emergence of hubness.} 
 Owing to the limited representational capacity induced by the polynomial volume growth of Euclidean space, a large number of entities are forced to crowd into a relatively small region of the embedding space. This geometric crowding gives rise to the hubness phenomenon \cite{hub1,hub2}: certain entities (i.e., hubs) frequently appear in the nearest neighbor sets of other entities.
 Such hubs tend to be repeatedly retrieved across diverse queries, leading to frequent co-occurrence across retrieval results and reducing retrieval diversity, as shown in Figure~\ref{fig:motivation}(b). As a consequence, Euclidean-based retrieval struggles to return query-specific and discriminative knowledge. 
In contrast, hyperbolic space exhibits exponential volume growth that naturally accommodates the expansion of hierarchical knowledge, alleviating geometric crowding. As illustrated in Figure~\ref{fig:motivation}(b), the number of high-frequency entities in hyperbolic retrieval is significantly reduced, demonstrating effective mitigation of the hubness problem.

To address these limitations, we propose a \underline{Hy}perbolic \underline{R}etrieval-\underline{A}ugmented \underline{G}eneration (HyRAG) framework to enhance the generalization capability of graph foundation models. Specifically, to model the tree-like hierarchies of the external knowledge base, we introduce the Hyperbolic Knowledge Indexing (HKI) module. This module pre-trains a hyperbolic semantic encoder via an unsupervised optimization paradigm, incorporating both the distance-based objective and the angular constraint to align with the geometric properties of hyperbolic entailment cones. To mitigate the loss of semantic granularity inherent in Euclidean-based retrieval, we develop a Multi-granularity Retrieval (MR) strategy, in which coarse-grained retrieval emphasizes diversity to capture global semantic anchors, whereas fine-grained retrieval focuses on capturing local semantic nuances.
Subsequently, tailored to the unique topology of graph data, we propose a Dual-path Fusion (DF) mechanism to effectively integrate knowledge at both the feature and structural levels. Extensive experiments across multiple graph benchmarks demonstrate that our method yields 
significant improvements in the inference accuracy of GFMs in the zero-shot setting. 

In summary, our main \textbf{contributions} are as follows: 1) We identify two inherent limitations of existing Euclidean-based RAG paradigms, namely the loss of semantic granularity and the emergence of hubness. 2) We propose HyRAG, the \textit{first} hyperbolic RAG framework tailored for graph foundation models, providing a geometry-aware redesign of the RAG pipeline encompassing knowledge indexing, retrieval, and fusion. 3) Extensive experiments on multiple graph benchmarks demonstrate that HyRAG consistently improves the zero-shot performance of GFMs.

\section{Related Work}

\subsection{Retrieval Augmented Generation}

Retrieval Augmented Generation retrieves relevant evidence from external knowledge bases at inference time and feeds it as additional context to large language models, improving factual consistency and reducing hallucinations \cite{rag4nlptask}. 
This paradigm has been widely adopted in knowledge-intensive NLP tasks, including open-domain question answering, fact verification, and multi-step reasoning \cite{DBLP:conf/eacl/IzacardG21,DBLP:journals/corr/abs-2401-15391}.
More recently, GraphRAG\cite{graphrag} and its variants \cite{lightrag,hipporag2,hypergraphrag} make inter-evidence relationships explicit by constructing document/entity graphs or indexing graphs, and incorporate graph traversal and community-level aggregation into retrieval, improving interpretability and multi-step evidence integration.

Beyond text-centric LLMs, recent efforts have started to bring RAG into GFMs to mitigate distribution shifts and enrich graph reasoning with external knowledge. In this line of work, retrieved evidence can come from auxiliary graphs, knowledge bases, or textual descriptions, and is injected into graph representations to support downstream tasks under few-shot or zero-shot settings \cite{ragraph,rag4gfm}. Specifically, RAGraph\cite{ragraph} builds a key–value toy-graph database from a resource graph, retrieves the most relevant toy graphs for a given query graph, and injects the retrieved representations and task signals into the query graph through a message-passing prompting mechanism.
RAG4GFM\cite{rag4gfm} constructs a multi-level index over a graph corpus as an external knowledge base, retrieves task-relevant subgraphs for a given query instance, and augments the input graph with the retrieved evidence to enable plug-and-play inference with a pre-trained GFM .
RAG-GFM\cite{rag-gfm} pioneers the integration of retrieval-augmented generation into the pre-training phase of graph foundation models, leveraging a dual-modal knowledge base to alleviate in-memory bottlenecks.
TarDGR\cite{tardgr} further introduces a task-aware retrieval framework that leverages task-specific supervision to guide the selection of relevant subgraphs in dynamic recommendation scenarios.

However, most existing methods perform representation learning and similarity matching in Euclidean space, which is suboptimal for knowledge with hierarchical or tree-like organization. To address this, we perform retrieval in hyperbolic space, which better preserves hierarchies and improves GFM retrieval and inference. 

\subsection{Hyperbolic Representation Learning}
Hyperbolic geometry has been widely adopted to model hierarchical and scale-free structures\cite{cone,du2022hakg,pan2021hyperbolic}. Early works, including the Poincaré ball model\cite{poincare} and the Lorentz\cite{lorentz} model, demonstrate that hyperbolic space can represent hierarchical relations with low distortion and high representational efficiency, providing a principled alternative to Euclidean embeddings.

Building upon these foundations, subsequent studies incorporate hyperbolic geometry into graph representation learning\cite{hypergcl,shang2024mixed,du2022hakg} and knowledge graph embeddings\cite{cone,pan2021hyperbolic} to capture hierarchical relations. Despite these advances, existing studies focus mainly on representation learning or supervised inference, while the application of hyperbolic geometry to RAG for GFMs has not been systematically investigated.
In contrast, this work investigates hyperbolic geometry from a retrieval-augmented reasoning perspective, and proposes a hyperbolic RAG framework tailored for GFMs.

\section{Preliminary}
\subsection{Notation}
Formally, we define the external knowledge base as $\mathcal{K}=\{(e_i,r_i,e_j$
$) |e_i,e_j\in \mathcal{E},r_i \in \mathcal{R}\}$, where $\mathcal{E}=\{e_1,\ldots,e_i,\ldots,e_N\}$ is the entity set and $\mathcal{R}=\{r_1,\ldots,r_i,\ldots,r_M\}$ denotes the relation set. Each triplet $(e_i, r_i, e_j)$ indicates that the head entity $e_i$ is connected to the tail entity $e_j$ via a relation $r_i$. Throughout this paper, we denote discrete entities and relations by lowercase italic letters (e.g., $e_i$ and $r_i$ ), and their continuous vector representations by boldface counterparts (e.g., $\mathbf{e}_i$ and $\mathbf{r}_i$ ).
\subsection{Hyperbolic Geometry}
Hyperbolic geometry is a non-Euclidean geometry with constant negative curvature. Unlike Euclidean space, where the volume of a metric ball grows polynomially with the radius, hyperbolic space exhibits exponential volume growth. This property makes it an ideal embedding space for tree-structured external knowledge bases.

\textbf{The Poincaré Ball Model.}
While hyperbolic space admits multiple isometric models, we adopt the Poincaré ball model\cite{poincare} for its conformality and geometric interpretability. In this work, we consider a $d$-dimensional Poincaré ball with fixed constant negative curvature $c=-1$, defined as:
\begin{equation}
    \mathbb{B}^d = \{ \mathbf{x} \in \mathbb{R}^d : \|\mathbf{x}\| < 1 \},
\end{equation}
where $\|\cdot\|$ denotes the standard Euclidean norm. The geometry of $\mathbb{B}^d$ is determined by the Riemannian metric $g^{\mathbb{B}}_{\mathbf{x}} = (\frac{2}{1-\|\mathbf{x}\|^2})^2 g^\mathbb{E}$, where $g^\mathbb{E}$ is the Euclidean metric. For notational simplicity, we omit the curvature parameter $c$ in all subsequent formulations.
On the Poincaré ball, the Möbius addition operator $\oplus$ between two points $\mathbf{x}, \mathbf{y} \in \mathbb{B}^d$ is defined as:
\begin{equation}
    \mathbf{x} \oplus \mathbf{y} = \frac{(1 + 2\langle \mathbf{x}, \mathbf{y} \rangle + \|\mathbf{y}\|^2)\mathbf{x} + (1 - \|\mathbf{x}\|^2)\mathbf{y}}{1 + 2\langle \mathbf{x}, \mathbf{y} \rangle + \|\mathbf{x}\|^2 \|\mathbf{y}\|^2}.
\end{equation}
Based on this definition, the hyperbolic distance between them is calculated as:
\begin{equation}
    d_{\mathbb{B}}(\mathbf{x}, \mathbf{y}) = 2\tanh^{-1}(\|-\mathbf{x} \oplus \mathbf{y}\|).
\end{equation}

\textbf{Tangent Space and Mappings.}
Given a point $\mathbf{x} \in \mathbb{B}^d$, the tangent space $T_\mathbf{x}\mathbb{B}$ is defined as the Euclidean vector space comprising all tangent vectors at $\mathbf{x}$. The exponential map, denoted as $\exp_\mathbf{x}(\cdot): T_\mathbf{x}\mathbb{B} \rightarrow \mathbb{B}^d$, facilitates the mapping of vectors from the tangent space to the manifold $\mathbb{B}^d$ as follows:
\begin{equation}
   \exp_{\mathbf{x}}(\mathbf{u}) = \mathbf{x} \oplus \tanh\left(\frac{\|\mathbf{u}\|}{1 - \|\mathbf{x}\|}\right)\frac{\mathbf{u}}{\|\mathbf{u}\|},
\end{equation}
while the logarithmic map $\log_{\mathbf{x}}(\cdot): \mathbb{B}^d \rightarrow T_{\mathbf{x}}\mathbb{B}$ serves as the inverse operation, projecting points from the manifold back to the tangent space:
\begin{equation}
    \log_{\mathbf{x}}(\mathbf{u}) = (1 - \|\mathbf{x}\|) \cdot \tanh^{-1}(\|-\mathbf{x} \oplus \mathbf{u}\|) \frac{-\mathbf{x} \oplus \mathbf{u}}{\|-\mathbf{x} \oplus \mathbf{u}\|}.
\end{equation}

\textbf{Hyperbolic Entailment Cones.}  
Hierarchical relations in the external knowledge base naturally induce a partial order, in which  coarse-grained concepts subsume more specific ones (e.g., artificial intelligence subsuming machine learning). Accordingly, to model such partial-order structures, we adopt hyperbolic entailment cones.
Each entity embedding $\mathbf{x} \in \mathbb{B}^d \setminus \{\mathbf{0}\}$ is associated with a hyperbolic entailment cone $\mathcal{C}_{\mathbf{x}}$, defined as
\begin{equation}
    \mathcal{C}_{\mathbf{x}} = \left\{ \mathbf{y} \in \mathbb{B}^d \;\middle|\; \angle_{\mathbf{x}}\mathbf{y} \leq \phi(\mathbf{x}) \right\}.
\end{equation}
This formulation encodes partial ordering via geometric transitivity:
if $\mathbf{y} \in \mathcal{C}_{\mathbf{x}}$, then $\mathcal{C}_{\mathbf{y}} \subseteq \mathcal{C}_{\mathbf{x}}$, ensuring that hierarchical entailment is preserved through cone containment.
$\angle_{\mathbf{x}}\mathbf{y}$ denotes the angle distance at $\mathbf{x}$ between the geodesic rays $\overrightarrow{\mathbf{o}\mathbf{x}}$ and $\overrightarrow{\mathbf{x}\mathbf{y}}$, where $\mathbf{o}$ denotes the origin of the Poincaré ball. Thus, $\angle_{\mathbf{x}}\mathbf{y}$ can be computed as
\begin{equation}
    \angle_{\mathbf{x}}\mathbf{y}
    =
    \cos^{-1}\!\left(
    \frac{
    \langle \mathbf{x}, \mathbf{y} \rangle (1+\|\mathbf{x}\|^2)
    -
    \|\mathbf{x}\|^2 (1+\|\mathbf{y}\|^2)
    }{
    \|\mathbf{x}\| \, \|\mathbf{x}-\mathbf{y}\|
    \sqrt{1+\|\mathbf{x}\|^2\|\mathbf{y}\|^2 - 2\langle \mathbf{x}, \mathbf{y} \rangle}
    }
    \right).
\end{equation}
The half-aperture $\phi(\mathbf{x})$ controls the width of the entailment cone and is defined as
\begin{equation}
    \phi(\mathbf{x}) = \sin^{-1}\!\left( K \frac{1 - \|\mathbf{x}\|^2}{\|\mathbf{x}\|} \right),
\end{equation}
where $K$ is a scaling hyper-parameter.
Under this formulation, the cone aperture is inversely related to the norm $\|\mathbf{x}\|$, naturally assigning wider cones to  coarse-grained concepts located near the origin and narrower cones to more fine-grained concepts.

\section{Methodology}
\begin{figure*}
  \centering
  \includegraphics[width=2\columnwidth]{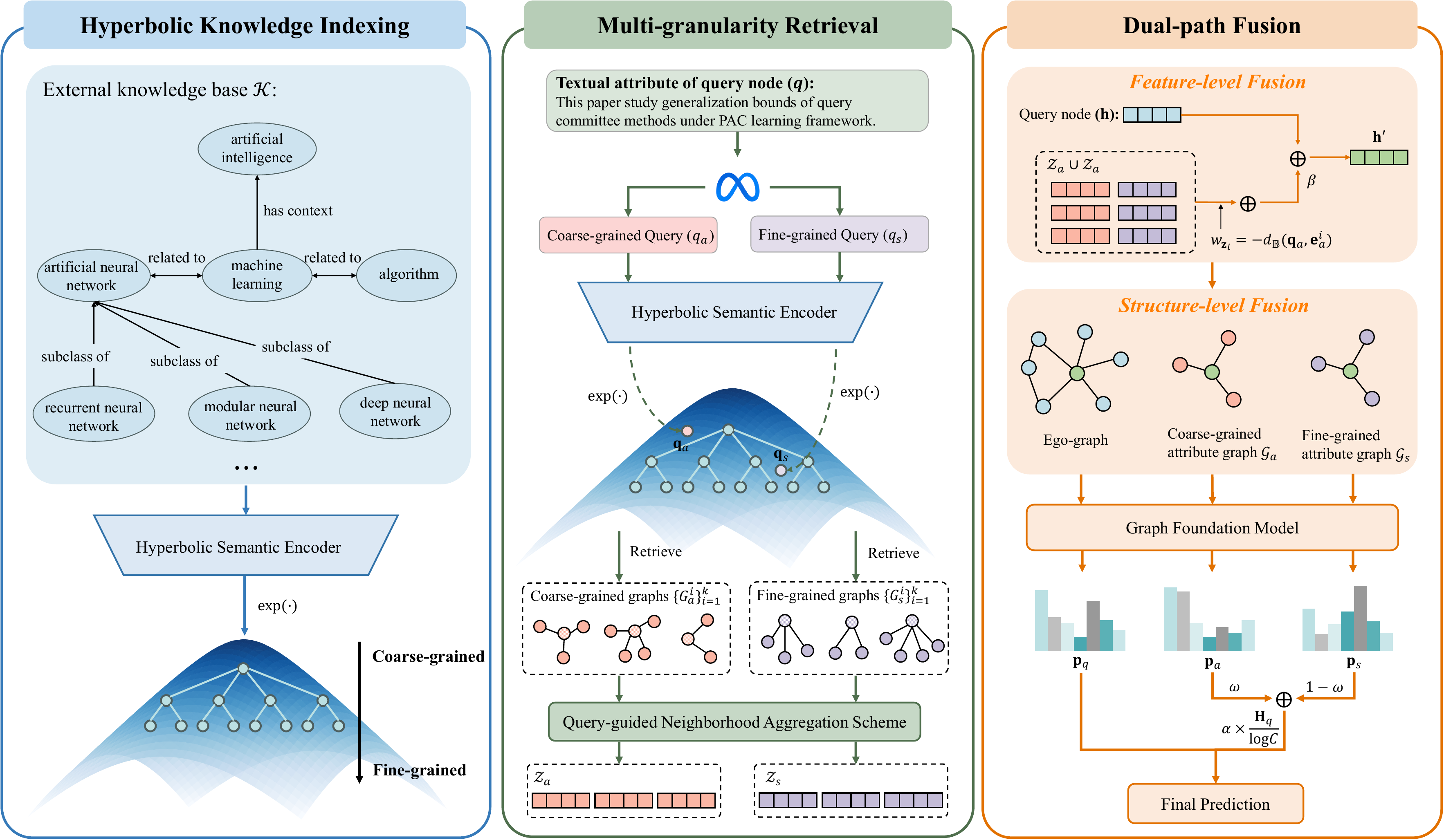}
\caption{Overall framework of our proposed HyRAG.}
  \label{fig:framework}
  \vspace{-0.3cm}
\end{figure*}
In this section, we present our proposed HyRAG, as depicted in Figure~\ref{fig:framework}.
HyRAG consists of three key modules: Hyperbolic Knowledge Indexing (HKI), Multi-granularity Retrieval (MR), and Dual-path Fusion (DF). 
These components work together to improve the generalization ability of GFMs by efficiently leveraging hierarchical knowledge structures via hyperbolic geometry.

\subsection{Hyperbolic Knowledge Indexing} \label{HKI}
To address the limitation of Euclidean embeddings in capturing the inherent hierarchy of the external knowledge base, we introduce the hyperbolic knowledge indexing module. This module aims to construct a hierarchy-aware hyperbolic representation space for the external knowledge base that serves as the geometric foundation for subsequent retrieval. Concretely, the module comprises two integral phases: hyperbolic projection and hierarchy-aware optimization.


\textbf{Hyperbolic projection.} This phase is designed to project the semantic features of the external knowledge base into a hyperbolic manifold. 

Specifically, we design a hyperbolic semantic encoder. To capture the semantics, the entity $e_i$ and relation $r_i$ in the external knowledge base are first encoded into Euclidean space using a pre-trained language model (e.g., SBERT), yielding $\mathbf{e}_i' \in \mathbb{R}^{d'}$ and $\mathbf{r}_i' \in \mathbb{R}^{d'}$.
Furthermore, a learnable mapping network $f_\Theta: \mathbb{R}^{d'} \rightarrow \mathbb{B}^d$ is learned to  transform the Euclidean embeddings into a hierarchy-aware latent space prior to hyperbolic projection.
The mapped representations are then projected onto the Poincaré ball model $\mathbb{B}^d$ via the exponential map, enabling the embeddings to reside in a negatively curved space where hierarchical structures can be naturally represented. This process can be formalized as:
\begin{equation}
    \mathbf{e}_i = \exp_{\mathbf{o}}(f_\Theta(\mathbf{e}_i')),\quad \mathbf{r}_i = \exp_{\mathbf{o}}(f_\Theta(\mathbf{r}_i')),
\end{equation}
where $\exp_{\mathbf{o}}(\cdot)$ denotes the exponential map at the origin.

\textbf{Hierarchy-aware optimization.} Inspired by \cite{cone}, we adopt a hierarchy-aware objective to jointly preserve relational semantics and hierarchical entailment in hyperbolic space.

Specifically, given a triplet $(e_i,r_i,e_j)\in\mathcal{K}$, the objective encourages the relation-conditioned representation of the head entity to be geometrically consistent with the corresponding tail entity in hyperbolic space, i.e.,
\begin{equation}
    \min_\Theta \; \mathbb{E}_{(e_i,r_i,e_j)\sim\mathcal{K}}
    \left[ d_{\mathbb{B}}(g(e_i,r_i), e_j) \right],
\end{equation}
where $g(e_i,r_i)$ denotes the relation-specific rotation function. For symmetric non-hierarchical relations (e.g., related to), we apply a direct rotation determined by the angle of the relation vector. 
For hierarchical relations (e.g., subclass-of), we adopt a rotation strategy under hyperbolic entailment cone constraints, ensuring that the transformed head entity remains within its original entailment cone after rotation along hyponym type relations. Consequently, the rotation function 
$g(\cdot)$ is formalized as:
\begin{equation}
\label{eq:rotation}
g(e_i, r_i) =
\begin{cases}
\exp_{\mathbf{o}}(\mathbf{R}(\theta_i) \log_{\mathbf{o}}(\mathbf{e}_i))  & \text{if } r_i \text{ is } \text{non-hierarchical},\\
\exp_{\mathbf{e}_i}\left(s_i \cdot \mathbf{R}\left(\theta_i \frac{\phi(\mathbf{e}_i)}{\pi}\right)\overline{\mathbf{e}}_i\right) & \text{otherwise.}
\end{cases}
\end{equation}
where $\mathbf{r}_i=(s_i,\theta_i)$, $\mathbf{R}(\cdot)$ is a rotation matrix:
\begin{equation}
    \mathbf{R}(\theta_i) = \begin{bmatrix} 
    \cos(\theta_i) & -\sin(\theta_i) \\ 
    \sin(\theta_i) & \cos(\theta_i) 
    \end{bmatrix},
\end{equation}
and $\phi(\mathbf{e}_i)$ is the half aperture of the cone $\mathbf{e}_i$. $\overline{\mathbf{e}}_i$ is the unit vector of $\mathbf{e}_i$ in the tangent space of $\mathbf{e}_i$:
\begin{align}
\overline{\mathbf{e}}_i = \widehat{\mathbf{e}}_i / \|\widehat{\mathbf{e}}_i\|, \quad \widehat{\mathbf{e}}_i = \log_{\mathbf{e}_i}\left(\frac{1 + \|\mathbf{e}_i\|}{2\|\mathbf{e}_i\|}\mathbf{e}_i\right) .
\end{align}
Notably, without loss of generality, we assume that the hierarchical relations in Eq.\ref{eq:rotation} correspond to hyponym type relations. Accordingly, to operationalize this minimization goal into a tractable training objective, we define the distance loss function using negative sampling:
\begin{equation}
    \mathcal{L}_{t} = -\log \sigma(d_{\mathbb{B}}(g(e_i,r_i), e_j)) - \mathbb{E}_{e_j' \sim \mathcal{T}} \left[ \log \sigma(-d_{\mathbb{B}}(g(e_i,r_i), e_j')) \right],
\end{equation}
where $(e_i, r_i, e_j)$ represents the positive sample, and the negative samples $(e_i, r_i, e_j')$ are constructed by corrupting the tail entity, where the negative entity $e_j'$ is uniformly sampled from a candidate set $\mathcal{T} \subseteq \mathcal{E} \setminus \{e_j\}$.

To explicitly model hierarchical containment, we further impose an angle constraint based on hyperbolic entailment cones. For instance, if the relation $r_i$ corresponds to a hyponym type relation, the hyperbolic entailment cone of the tail entity $e_j$ is required to be geometrically contained within that of the head entity $e_i$. Thus, we adopt an angle loss function:
\begin{equation}
    \mathcal{L}_{a} = \max(0, \angle_{\mathbf{e}_i}\mathbf{e}_j - \phi(\mathbf{e}_i)).
\end{equation}
However, optimizing only the distance loss and the angle loss may cause embeddings to drift toward the boundary of the Poincaré ball, leading to degenerate hierarchical representations.
To mitigate this effect, we penalize embeddings whose radial norms exceed a predefined threshold $\epsilon$, thereby stabilizing the geometry and preserving effective representational capacity:
\begin{equation}
\mathcal{L}_{\text{reg}} =  \max(0, \|\mathbf{e}_i\| - \epsilon)^2 + \max(0, \|\mathbf{e}_j\| - \epsilon)^2.
\end{equation}
Thus, the final objective is given by:
\begin{equation}
    \mathcal{L} = \mathcal{L}_{t} + \lambda_a \mathcal{L}_{a}+\lambda_{\text{reg}} \mathcal{L}_{\text{reg}},
\end{equation}
where $\lambda_a$ and $\lambda_{\text{reg}}$ are hyper-parameters.
By optimizing the mapping network $f_\Theta$ through this objective function, we obtain the hyperbolic index for entities in the external knowledge base, denoted as $\mathcal{H}=\{\mathbf{e}_1,\ldots,\mathbf{e}_i,\ldots,\mathbf{e}_N\}$, where each $\mathbf{e}_i \in \mathbb{B}^d$ resides in the Poincaré ball. This embedding space jointly captures rich textual semantics and the underlying hierarchical structure of the knowledge base, thereby mitigating the hubness problem.

\subsection{Multi-granularity Retrieval}
As discussed in Section~\ref{intro}, coarse-grained and fine-grained concepts provide distinct perspectives essential for graph tasks. To this end, we propose the multi-granularity retrieval mechanism designed to retrieve complementary knowledge at varying granularities.

Rather than imposing hard geometric constraints (e.g., explicit radius thresholds), we leverage the large language model (LLM) to semantically reformulate the original textual attribute $q$ of the query node into two distinct variants: 1) a coarse-grained query $q_a$ that summarizes the core theme, and 2) a fine-grained query $q_s$ that captures fine-grained details. By leveraging the pre-trained mapping network $f_\Theta(\cdot)$, these reconstructed queries are naturally embedded into the hyperbolic space at positions commensurate with their semantic granularities. Consequently, they are subsequently employed to retrieve relevant knowledge from the hyperbolic index $\mathcal{H}$ through coarse-grained and fine-grained knowledge retrieval processes.

\textbf{Coarse-grained knowledge retrieval.}
Given a coarse-grained query $q_a$, it is first processed by the pre-trained hyperbolic semantic encoder, which comprises the language model and the mapping network $f_\Theta$, and subsequently projected into the hyperbolic space via the exponential map:
\begin{equation}
    \mathbf{q}_a = \exp_\mathbf{o}(f_\Theta(\text{LM}(q_a))),
\end{equation}
where $\mathbf{q}_a$ is the resulting hyperbolic embedding of $q_a$. 
Based on this representation, the top-$k$ entities with the smallest hyperbolic distances to $\mathbf{q}_a$ are retrieved from the hyperbolic index $\mathcal{H}$:
\begin{equation}
\mathcal{E}_a^\text{top} = \topk_{\mathbf{e}_i\in \mathcal{H}} d_\mathbb{B}(\mathbf{q}_a, \mathbf{e}_i),
\label{eq:a_topk}
\end{equation}
where $\mathcal{E}_a^\text{top}$ denotes the selected set of top-$k$ coarse-grained entities. 

To further enrich the retrieved context with structural information, we expand each entity $e_a \in \mathcal{E}_a^\text{top}$ by incorporating its neighborhood entities from the external knowledge base. Let $\mathcal{N}(e_a)$ denote the neighborhood set of $e_a$. Directly aggregating all neighboring entities may introduce substantial redundancy, as many neighbors tend to be semantically similar to each other. To balance relevance to the query and diversity among the selected neighbors, we apply the Maximal Marginal Relevance (MMR) criterion~\cite{mmr} to re-rank candidate entities. Specifically, at each selection step, the candidate entity $e_a^{*} \in \mathcal{N}(e_a)\setminus \mathcal{S}(e_a)$ is chosen by maximizing a trade-off between its relevance to the query $q_a$ and its dissimilarity to the already selected set $\mathcal{S}(e_a)$:
\begin{equation}
    e_a^{*} = \argmax_{e_i \in \mathcal{N}(e_a)\setminus \mathcal{S}(e_a) } \{-\gamma \cdot d_\mathbb{B}(\mathbf{e}_i, \mathbf{q}_a) + (1-\gamma) \cdot \min_{\mathbf{e}_j \in \mathcal{S}(e_a)} d_\mathbb{B}(\mathbf{e}_i, \mathbf{e}_j)\},
\end{equation}
where $\gamma \in [0,1]$ controls the trade-off between relevance and diversity. The selection process is iteratively performed until a predefined number of neighboring entities has been selected. 

As a result, the coarse-grained knowledge retrieval stage produces a set of 
$k$ graphs, denoted as $\{G_a^i\}_{i=1}^k$. Each graph $G_a^i$ consists of a central entity $e_a^i \in \mathcal{E}_a^{\text{top}}$ together with its neighboring entities selected via the MMR criterion.

\textbf{Fine-grained knowledge retrieval.}
For fine-grained knowledge retrieval, the textual query $q_s$ is encoded into the hyperbolic space using the same hyperbolic semantic encoder and exponential map, yielding its hyperbolic embedding $\mathbf{q}_s$:
\begin{equation}
    \mathbf{q}_s = \exp_\mathbf{o}(f_\Theta(\text{LM}(q_s))).
\end{equation}
Subsequently, the top-$k$ entities closest to $\mathbf{q}_s$ in the hyperbolic space are selected based on the hyperbolic distance:
\begin{equation}
\mathcal{E}_s^\text{top} = \topk_{\mathbf{e}_i\in \mathcal{H}} d_\mathbb{B}(\mathbf{q}_s, \mathbf{e}_i),
\label{eq:s_topk}
\end{equation}
where $\mathcal{E}_s^\text{top}$ is the selected set of top-$k$ fine-grained entities. 

Following the same hyperbolic retrieval paradigm as in the coarse-grained stage, neighborhood expansion in the fine-grained knowledge retrieval stage is also guided by geometric proximity, but employs a different selection criterion to emphasize fine-grained specificity.
Instead of applying diversity-based re-ranking, we first construct a candidate pool $\mathcal{C}_s$ by retaining entities whose hyperbolic distances to $\mathbf{q}_s$ fall within a predefined proximity range, resulting in a relatively large set of candidate entities. 
Motivated by the concept of hyperbolic entailment cones, we further impose an angular constraint to capture hierarchical and directional consistency. Specifically, for each $e_s \in \mathcal{E}_s^\text{top}$, an angular violation score is defined for each candidate entity according to the following formulation:
\begin{equation}
    \text{score}(e_i) = \max(0, \angle_{\mathbf{e}_s}\mathbf{e}_i - \phi(\mathbf{e}_s)), \;\forall e_i \in \mathcal{C}_s.
\end{equation}
Under the definition of hyperbolic entailment cones, a smaller value of $\text{score}(e_i)$ indicates that the entity $e_i$ is more likely to be a hyponym of $e_s$. Accordingly, candidate entities are ranked based on this score, and the top-$k'$ entities with the smallest scores are selected as neighborhood entities for $e_s$, thereby revealing fine-grained semantic information.

The fine-grained knowledge retrieval stage ultimately produces 
$k$ graphs, denoted as $\{G_s^i\}_{i=1}^k$. Each graph is centered on a retrieved fine-grained entity $e_s^i \in \mathcal{E}_s^\text{top}$ and augmented with neighborhood entities selected under the angular constraint.

Finally, the retrieved coarse-grained and fine-grained graphs are summarized into compact graph-level representations.
Given the retrieved coarse-grained graph $G_a^i$ induced by $e_a^i$, its graph-level representation is obtained through the query-guided neighborhood aggregation scheme:
\begin{equation}
    \mathbf{z}_a^i = \mathbf{e}_a^i + \sum_{e_i \in \mathcal{N}^*(e_a^i)} \frac{\exp(\mathbf{w}_{e_i}/\tau)}{\sum_{e_j \in \mathcal{N}^*(e_a^i)} \exp(\mathbf{w}_{e_j}/\tau)} \mathbf{e}_i, 
\end{equation}
where $\mathcal{N}^*(e_a^i)$ denotes the neighborhood of entity $e_a^i$ in graph $G_a^i$, and $\mathbf{w}_{e_i} = -d_\mathbb{B}(q_a,e_i)$ is the relevance score based on hyperbolic distance, while $\tau$ is the temperature coefficient. Following this procedure, a coarse-grained knowledge set $\mathcal{Z}_a =\{\mathbf{z}_a^1,\ldots,\mathbf{z}_a^k\}$ and a fine-grained knowledge set $\mathcal{Z}_s =\{\mathbf{z}_s^1,\ldots,\mathbf{z}_s^k\}$ are obtained, where $\mathbf{z}_a^i$ denotes the representation of $G_a^i$ and $z_s^i$ denotes the representation of $G_s^i$.
\subsection{Dual-path Fusion}
Motivated by the fact that graph data is defined by both node attributes and topological connections, our proposed dual-path fusion mechanism consists of feature-level fusion and structure-level fusion. This design allows the model to simultaneously augment local node semantics and enhance global structures.
\begin{table*}[t]
 \caption{Accuracy (\%) for the node classification task in the zero-shot setting. We report the average accuracy in the last column, and the best performance is highlighted in bold. ``GraphCLIP+Vanilla'' denotes the original implementation of GraphCLIP.}
\renewcommand\arraystretch{1.3}
\adjustbox{max width=0.86\textwidth}{\begin{tabular}{l|ccccccccc}
\toprule
Methods &  Cora & CiteSeer & WikiCS & Instagram & Ele-Photo & Ele-Computers & Books-History & AVG.\\

\midrule
DGI \cite{dgi} & 24.03$\pm$1.40 & 18.71$\pm$1.22 & 18.86$\pm$0.25 & 61.42$\pm$1.12 & 13.96$\pm$0.17 & 27.12$\pm$0.03 & 15.77$\pm$0.02 &25.70\\
GRACE \cite{grace} & 13.69$\pm$1.27 & 22.88$\pm$1.49 & 16.07$\pm$0.32 & 62.23$\pm$0.93 & 10.16$\pm$0.13 & 10.94$\pm$0.12 & 32.39$\pm$0.11 &24.05\\
BGRL \cite{bgrl} & 31.99$\pm$1.06 & 26.50$\pm$1.22 & 18.35$\pm$0.22 & 61.45$\pm$0.82 & 5.21$\pm$0.22 & 24.12$\pm$0.22 & 16.28$\pm$0.35 &26.41\\
GraphMAE \cite{graphmae} &23.25$\pm$1.07 & 20.75$\pm$0.88 & 12.14$\pm$0.20 & 62.39$\pm$0.84 & 12.53$\pm$0.08 & 8.36$\pm$0.06 & 21.76$\pm$0.17 & 23.03\\
G2P2 \cite{g2p2} &41.51$\pm$0.78 & 51.02$\pm$0.62 & 31.92$\pm$0.15 & 52.87$\pm$0.78 & 22.21$\pm$0.12 & 32.52$\pm$0.13 & 26.18$\pm$0.25 &36.89\\
\midrule
GraphGPT  & 23.25$\pm$1.45 & 18.04$\pm$1.45 & 6.30$\pm$0.26 & 45.12$\pm$1.16 & 7.62$\pm$0.22 & 29.71$\pm$0.83 & 15.92$\pm$0.14 &20.85\\
LLaGA  &21.44$\pm$0.65 & 16.07$\pm$1.15 & 2.65$\pm$0.72 & 41.12$\pm$0.94 & 6.50$\pm$0.53 & 23.10$\pm$0.33 & 11.17$\pm$0.58 &17.44\\
OFA & 37.25$\pm$1.38 & 29.64$\pm$0.19 & 45.52$\pm$1.06 & 32.71$\pm$0.16 & 33.03$\pm$0.64 & 22.09$\pm$0.39 & 16.87$\pm$0.93 &30.87\\
ZeroG  &  62.32$\pm$1.91 & 52.55$\pm$1.23 & 54.93$\pm$0.06 & 48.97$\pm$0.78 & 45.12$\pm$0.65 & 56.20$\pm$0.35 & 40.74$\pm$0.65 &51.40\\
UniGraph & 69.53$\pm$0.00 & - & 43.45$\pm$0.00 & - & - & - & - &56.49\\
\midrule

\multicolumn{9}{l}{\textbf{GraphCLIP+}}\\
\hline
Vanilla & $67.31\pm1.76$ & $63.13 \pm 1.13$ & $70.19\pm0.10$ & $64.05\pm0.34$ & $53.40\pm0.64$ & $62.04\pm0.21$ & $53.88\pm0.35$ &62.00\\
RAGRAPH & $68.63\pm0.75$ & $67.19\pm0.94$ & $71.40\pm0.17$ & $63.79\pm0.42$ & $54.39\pm0.35$ & $64.26\pm0.26$ & $53.70\pm0.22$ &63.34\\
\hline
HyRAG & $\mathbf{68.94\pm0.57}$ & $\mathbf{67.24\pm0.80}$ & $\mathbf{72.66\pm0.73}$ & $\mathbf{64.05\pm0.31}$ & $\mathbf{56.05\pm0.55}$ & $\mathbf{65.04\pm0.28}$ & $\mathbf{55.14\pm0.35}$ & \textbf{64.16}\\
\bottomrule
\end{tabular}}

 \label{tab:main}
\end{table*}

\textbf{Feature-level fusion.} 
Feature-level fusion is achieved by aggregating both coarse-grained and fine-grained knowledge into the initial representation of the query node.
The fusion weights are computed based on the hyperbolic distances between the query node representation and the central entity representations of the retrieved graphs.
Without loss of generality, we assume that $\mathbf{z}_i \in \mathcal{Z}_a$. Then, the corresponding weight $\mathbf{w}_{\mathbf{z}_i}$ is computed as the hyperbolic distance between the query embedding $\mathbf{q}_a$ and the representation of its central entity $\mathbf{e}_a^i$, i.e.,
\begin{equation}
    \mathbf{w}_{\mathbf{z}_i} = -d_\mathbb{B}(\mathbf{q}_a,\mathbf{e}_a^i).
\end{equation}
Based on this, the feature-level fusion can be formalized as:
\begin{equation}
    \mathbf{h}' = \mathbf{h} + \beta \sum_{\mathbf{z}_i \in \mathcal{Z}_a \cup \mathcal{Z}_s} \frac{\exp(\mathbf{w}_{\mathbf{z}_i}/\tau)}{\sum_{\mathbf{z}_j\in \mathcal{Z}_a \cup \mathcal{Z}_s} \exp(\mathbf{w}_{\mathbf{z}_j}/\tau)} \mathbf{z}_i,
\end{equation}
where $\mathbf{h}$ represents the initial representation of the query node, and $\beta$ controls the contribution of knowledge injection.

\textbf{Structure-level fusion.}
GFMs typically perform downstream prediction based on the ego-graph of the query node, where edges represent intrinsic interactions between nodes, such as citation links in citation networks. However, the retrieved external knowledge does not naturally conform to the ego-graph topology. Directly connecting retrieved central entities to the query node within the ego-graph would introduce edges with heterogeneous semantics, resulting in a mixed graph structure that may degrade the quality of graph encoding.

To avoid structural semantic inconsistency, we construct two auxiliary graphs, namely the coarse-grained attribute graph $\mathcal{G}_a$ and the fine-grained attribute graph $\mathcal{G}_s$. In both graphs, the retrieved central entities are directly connected to the query node.
Then, the graph encoder in the GFM is employed to independently encode the ego-graph of the query node, the coarse-grained attribute graph, and the fine-grained attribute graph, yielding the corresponding prediction logits $\mathbf{p}_{q}, \mathbf{p}_{a}, \mathbf{p}_{s}$. To effectively integrate these multi-view predictions, we further propose an uncertainty-gated knowledge injection mechanism. 

Specifically, we first fuse the coarse-grained and fine-grained predictions into a unified knowledge prediction $\mathbf{p}_\text{know}$ as follows:
\begin{equation}
    \mathbf{p}_\text{know} = \omega \times \mathbf{p}_{a} + (1-\omega) \times \mathbf{p}_{s},
\end{equation}
where the weight $\omega$ is determined by the confidence score, defined as the exponential of the inverse entropy:
\begin{equation}
    \omega = \frac{\exp\left(\frac{1}{1 + \mathbf{H}_a}\right)}{\exp\left(\frac{1}{1 + \mathbf{H}_a}\right) + \exp\left(\frac{1}{1 + \mathbf{H}_s}\right)},
\end{equation}
where $\mathbf{H}_a$ and $\mathbf{H}_s$ are the entropies of $\mathbf{p}_a$ and $\mathbf{p}_s$. Crucially, this mechanism serves to assign higher fusion weights to more credible knowledge sources, ensuring that the unified knowledge prediction relies more heavily on the granularity with lower uncertainty.
Finally, we inject the unified knowledge prediction $\mathbf{p}_\text{know}$ into the original prediction. 
Crucially, this injection is gated by the uncertainty of the original prediction and can be formalized as:
\begin{equation}
    \mathbf{p}_\text{final} = \mathbf{p}_{q} + \alpha \times \frac{\mathbf{H}_q}{\log C} \times \mathbf{p}_\text{know},
\end{equation}
where $\frac{\mathbf{H}_q}{\log C}$ denotes the normalized entropy of the original prediction, $C$ is the number of categories, and $\alpha$ is a hyper-parameter. This integration mechanism acts as an adaptive switch: when the ego-graph is informative (i.e., low entropy), the model preserves its prediction to avoid noise; when the ego-graph is ambiguous (i.e., high entropy), the model actively leverages the unified knowledge prediction $\mathbf{p}_\text{know}$ to rectify the prediction.



\section{Experiment}
\subsection{Experimental Settings}
\textbf{Datasets.} 
We conduct experiments on seven widely used graph datasets spanning four domains. Specifically, Cora \cite{cora} and CiteSeer \cite{citeseer} are from the Academic domain; Ele-Photo \cite{cstag}, Ele-Computers \cite{cstag}, and Books-History \cite{cstag} are from the E-commerce domain; WikiCS \cite{wikics} is from the Wikipedia domain; and Instagram \cite{instagram} is from the Social domain. More details about the datasets are provided in Appendix~\ref{sec:datasets}.

\textbf{Baselines.} We conduct comparisons against 12 baseline methods, categorized as follows: (i) self-supervised graph methods, including DGI \cite{dgi}, GRACE \cite{grace}, BGRL \cite{bgrl}, GraphMAE \cite{graphmae}, and G2P2 \cite{g2p2};
(ii) state-of-the-art graph foundation models, including GraphGPT \cite{graphgpt}, LLaGA \cite{llaga}, OFA \cite{ofa}, ZeroG \cite{zerog}, UniGraph\cite{unigraph}, GraphCLIP \cite{graphclip}, and AnyGraph \cite{anygraph}; (iii) RAG-based approach specifically designed for graph tasks, including RAGRAPH\cite{ragraph}.

\textbf{Experimental Details.} 
We conduct experiments under a zero-shot setting without updating the parameters of GFMs. The objective function of the HKI module is used solely to pre-train the mapping network 
$f_\Theta(\cdot)$, which is then frozen at inference time of GFMs.
The external knowledge base used for retrieval is derived from a subset of the commonsense knowledge graph (CSKG)\cite{cskg}, comprising ConceptNet\cite{cskg}, WordNet\cite{cskg}, and Wikidata-CS\cite{cskg}\footnote{Notably, Wikidata-CS refers to the commonsense knowledge in Wikidata, while the WikiCS dataset denotes the computer science branch within Wikidata.}. For all experimental results, we conduct three independent runs and report the mean and standard deviation. More experimental details can be found in Appendix~\ref{sec:imp}.


\subsection{Generalization Performance}
\begin{figure}
  \centering
  \includegraphics[width=1\columnwidth]{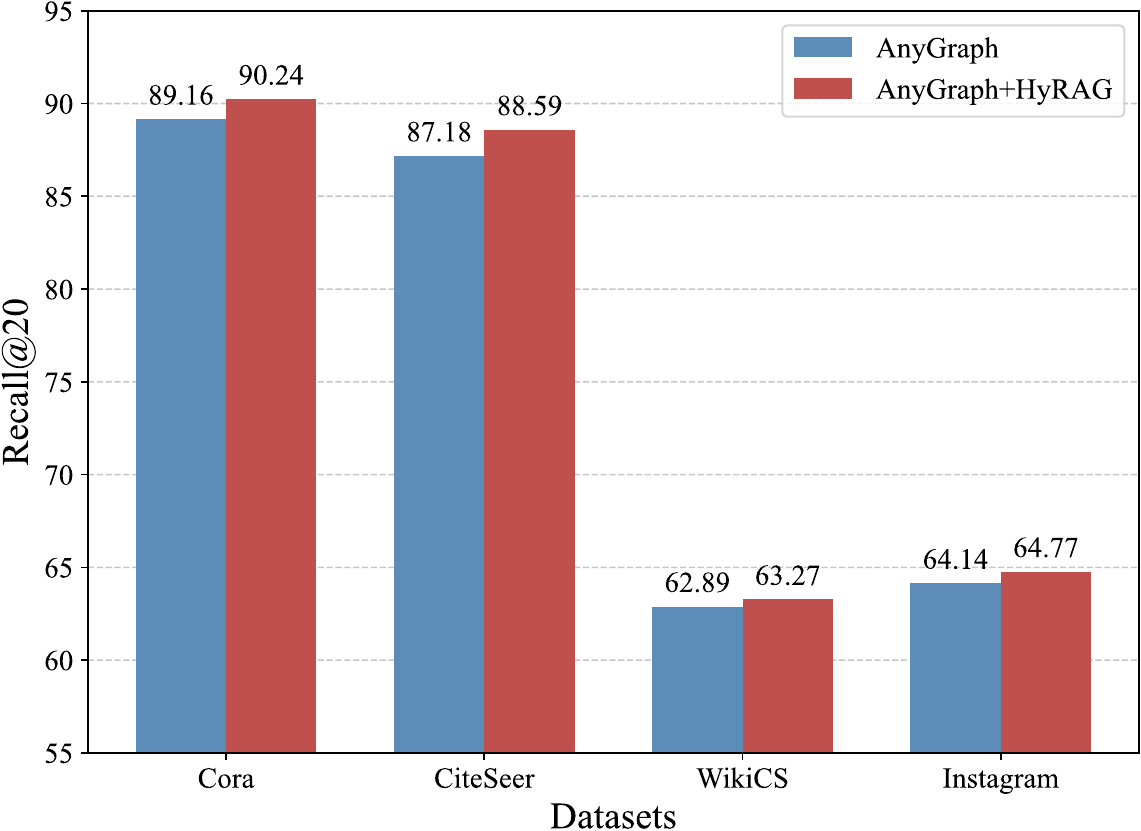}
\caption{Accuracy (\%) for the link prediction task in the zero-shot setting.
}
  \label{fig:link}
\end{figure}

\textbf{Performance on the node classification task.} To evaluate the effectiveness of HyRAG in enhancing the generalization performance of GFMs, we integrate HyRAG with the pre-trained graph foundation model GraphCLIP as the backbone. Table~\ref{tab:main} reports the zero-shot node classification accuracy across seven unseen datasets from diverse domains. Overall, HyRAG consistently achieves the best performance across all datasets. Compared to state-of-the-art self-supervised graph representation learning methods and graph foundation models, HyRAG yields an average improvement of up to 2.02\%. This result empirically supports our analysis in Section~\ref{intro}, indicating that the parameterized knowledge encoded within model weights is insufficient to handle distribution shifts in the zero-shot setting. By dynamically retrieving external knowledge, HyRAG effectively enhances the generalization of GFMs. Furthermore, compared to the Euclidean-space retrieval method RAGRAPH, HyRAG achieves an average improvement of 0.82\%. This gain highlights the advantage of hyperbolic geometry over Euclidean geometry for knowledge retrieval, demonstrating the effectiveness of our hyperbolic RAG framework.

\textbf{Performance on the link prediction task.}
We additionally conduct experiments on the link prediction task, and the results are shown in Figure \ref{fig:link}. HyRAG consistently improves the performance of the backbone model AnyGraph across all datasets. In particular, HyRAG achieves noticeable gains on Cora and CiteSeer, improving the accuracy by 1.08\% and 1.41\%, respectively. We attribute these improvements to the multi-granularity retrieval mechanism in hyperbolic space, which better preserves semantic dependencies and alleviates the hubness issue of Euclidean retrieval. Moreover, consistent gains on WikiCS and Instagram further demonstrate that the retrieved knowledge provides complementary structural and semantic evidence for link prediction. Overall, these results verify that HyRAG generalizes effectively beyond node classification and consistently enhances the relational reasoning capability of GFMs in zero-shot settings.

\begin{table}
\caption{The results of the ablation experiment.}
\vspace{-0.2cm}
\centering
\adjustbox{max width=\columnwidth}{
\begin{tabular}{ccccc|ccc}
\toprule
  \multirow{2}{*}{HKI}&\multicolumn{2}{c}{MR}&\multicolumn{2}{c|}{DF}  & \multirow{2}{*}{WikiCS} & \multirow{2}{*}{Ele-Photo}& \multirow{2}{*}{Books-History} \\ \cmidrule(lr){2-3}  \cmidrule(lr){4-5}
  & CR &FR & FF & SF&&&\\ 
\midrule 
$\times$ & $\times$  & $\times$ & $\times$ & $\times$ & $70.19\pm0.10$ & $53.40\pm0.64$  &  $53.88\pm0.35$ \\
\midrule
$\times$ & \checkmark &\checkmark & \checkmark &\checkmark &$70.94\pm0.93$ & $53.60\pm0.35$ & $53.35\pm0.10$ \\
\checkmark & $\times$ & \checkmark & \checkmark &\checkmark &$72.26\pm1.08$ & $55.00\pm0.38$ & $55.10\pm0.45$ \\
 \checkmark & \checkmark & $\times$ &\checkmark &\checkmark &$71.19\pm0.90$ & $55.96\pm0.52$ & $54.36\pm0.14$ \\
\checkmark & \checkmark & \checkmark & $\times$ &\checkmark &$71.85\pm0.74$ & $54.37\pm0.27$ & $53.84\pm0.31$  \\
\checkmark & \checkmark & \checkmark &\checkmark & $\times$  &$72.55\pm0.66$ & $55.79\pm0.46$ & $54.93\pm0.33$  \\
\checkmark & \checkmark & \checkmark &\checkmark & \checkmark  &$\mathbf{72.66\pm0.73}$ & $\mathbf{56.05\pm0.55}$ & $\mathbf{55.14\pm0.35}$  \\
\bottomrule
\end{tabular}
\label{tab:ablation}}
\end{table}
\subsection{Ablation Study}
We perform ablation experiments to assess the contribution of each component in HyRAG, and present the results in Table~\ref{tab:ablation}. The first row shows the performance of the baseline method, i.e., GraphCLIP. The last row reports the performance of HyRAG. 

\subsubsection{The Effectiveness of Hyperbolic Knowledge Indexing} 
The second row reports the results where the hyperbolic knowledge indexing module is replaced by the standard Euclidean index. Notably, this substitution results in the largest performance degradation among all ablation variants, with an average decline of 1.99\%. This significant drop empirically validates the superiority of hyperbolic geometry over Euclidean space in preserving the hierarchical structure of the external knowledge base, thereby confirming the necessity of our proposed hyperbolic RAG framework.
\begin{figure}
  \centering
  \includegraphics[width=1\columnwidth]{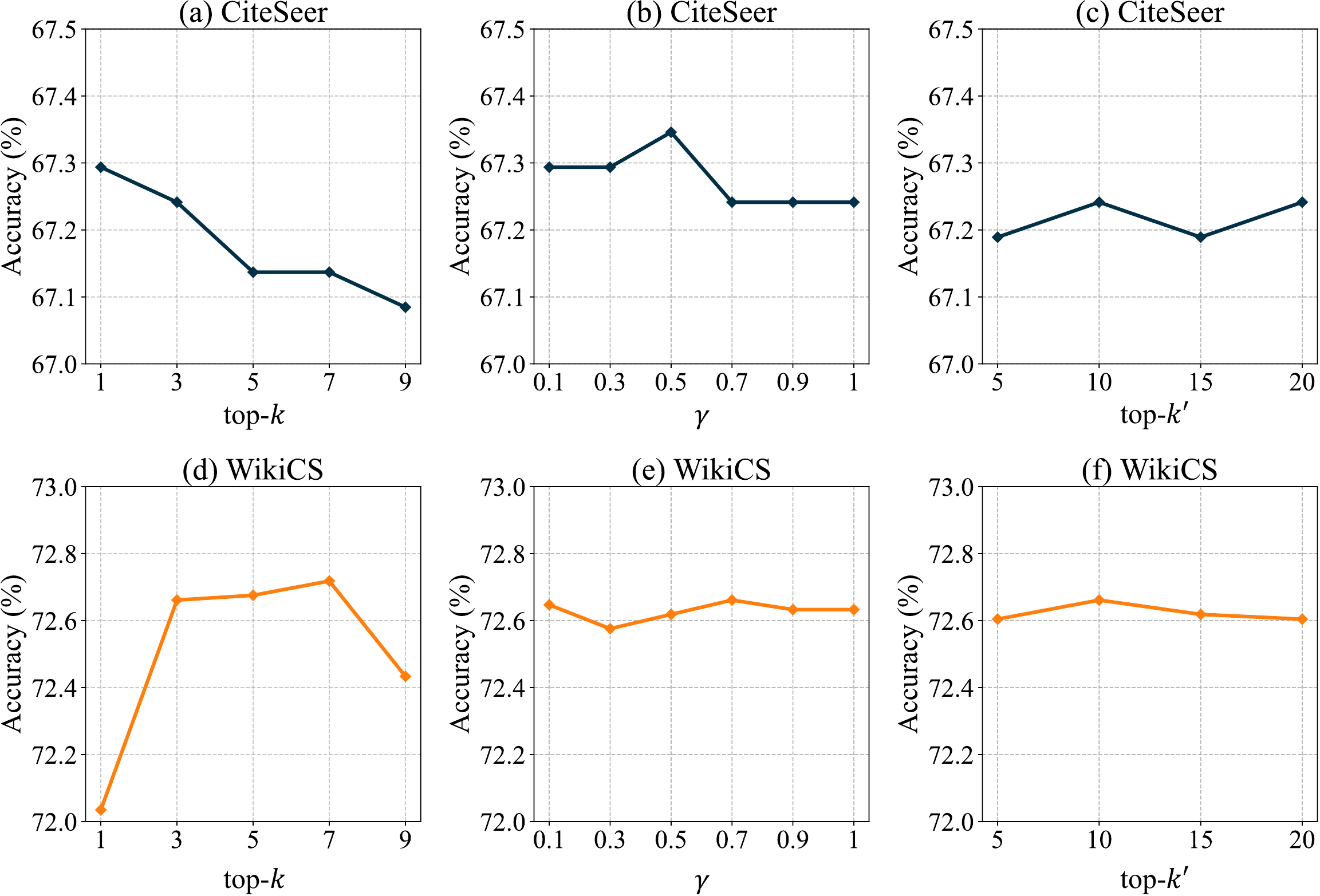}
\caption{ Hyper-parameter sensitivity analysis of the multi-granularity retrieval module. 
}
  \label{fig:hyper}
\end{figure}

\subsubsection{The Effectiveness of Multi-granularity Retrieval}
The third and fourth rows show the results of removing coarse-grained knowledge retrieval (CR) and fine-grained knowledge retrieval (FR), respectively. Overall, removing either component leads to a noticeable performance degradation, indicating that both components are indispensable for downstream graph tasks. Notably, for the Ele-Photo dataset, the performance deterioration is more pronounced when removing CR. We hypothesize that this is because node attributes in Ele-Photo are derived from user reviews, which inevitably incorporate substantial noise. The coarse-grained retrieval mitigates this issue by capturing representative global semantic anchors, thereby enhancing the robustness of GFMs against such noise interference.
For the WikiCS and Books-History datasets, removing FR leads to a larger performance drop.
This is likely because these datasets contain categories with subtle distinctions and the fine-grained knowledge proves critical by providing local semantic nuances for precise discrimination. In summary, these findings confirm that coarse-grained and fine-grained knowledge benefit GFMs from distinct yet synergistic perspectives.

\subsubsection{The Effectiveness of Dual-path Fusion} 
The ablation results in the fifth and sixth rows highlight the distinct contributions of our dual-path fusion strategy. Notably, the model exhibits higher sensitivity to the removal of feature-level fusion (FF). This suggests that, in a zero-shot setting, leveraging retrieved multi-granularity knowledge to enhance the initial node features is the fundamental driving force for achieving generalization. In parallel, structure-level fusion (SF) facilitates message passing over the graph, acting as a complementary mechanism to further boost performance on top of the enriched semantic features.
\begin{figure}
  \centering
  \includegraphics[width=1\columnwidth]{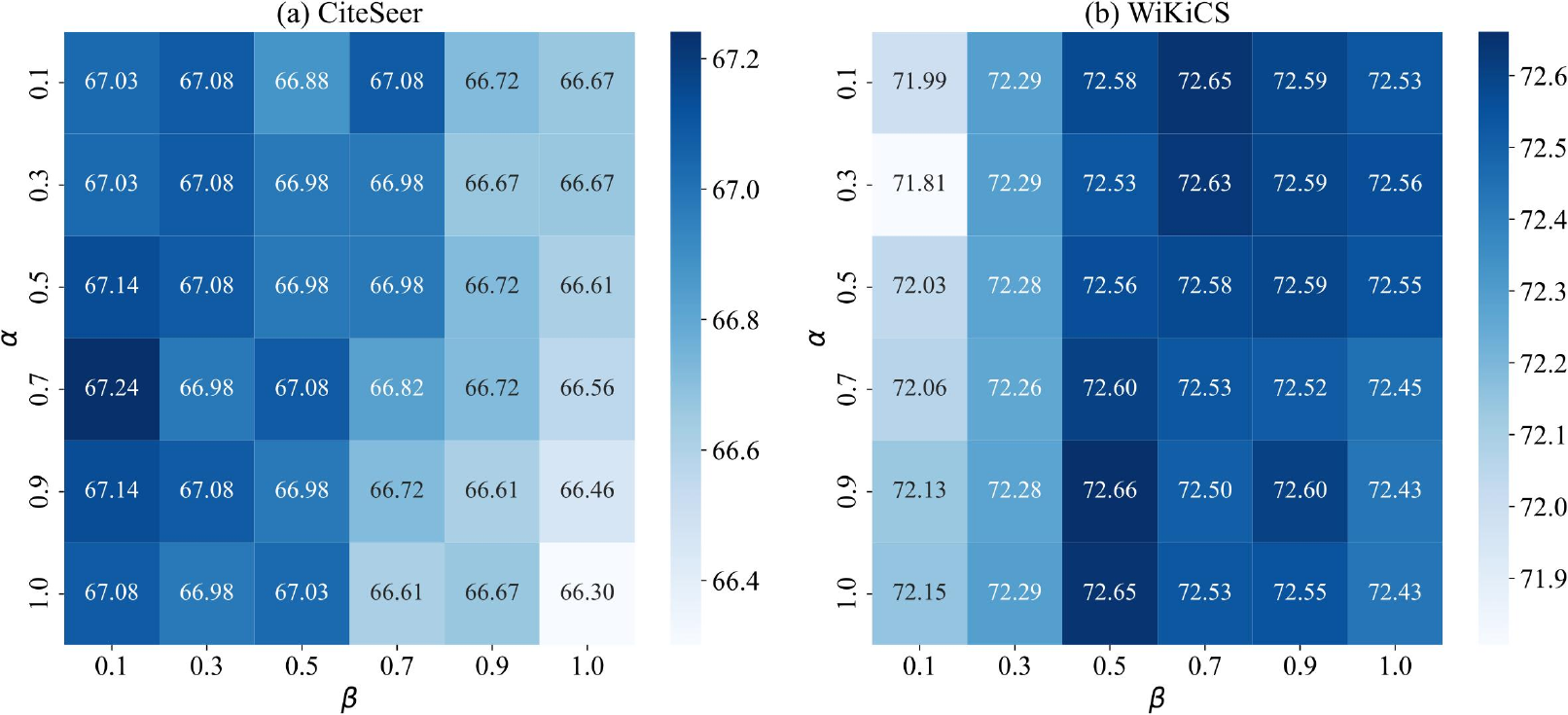}
\caption{ Hyper-parameter sensitivity analysis of $\beta$ and $\alpha$.
}
  \label{fig:hotmap}
\end{figure}

\subsection{Hyper-parameter Study}
\subsubsection{Hyper-parameters in Multi-granularity Retrieval}
We analyze the impact of three hyper-parameters in the multi-granularity retrieval module on the CiteSeer and WikiCS datasets.

Hyper-parameter $k$ controls the amount of information retrieved in the multi-granularity retrieval module. On CiteSeer (Figure~\ref{fig:hyper}(a)), the accuracy decreases as $k$ increases, indicating that CiteSeer requires precise retrieval and that incorporating excessive information introduces irrelevant noise. In contrast, WikiCS (Figure~\ref{fig:hyper}(d)) benefits from a larger $k$ (reaching its peak at $k=7$), which may be due to its fine-grained categories requiring broader semantic coverage to distinguish similar categories. 

Hyper-parameter $\gamma$ controls the trade-off between relevance and diversity in the MMR algorithm. Figure~\ref{fig:hyper}(b) reveals a performance peak at $\gamma=0.5$ for CiteSeer, indicating an optimal trade-off between relevance and diversity is necessary for this dataset. In contrast, WikiCS (Figure~\ref{fig:hyper}(e)) shows remarkable robustness to $\gamma$.

The hyper-parameter $k'$ governs the number of retrieved neighborhood entities in fine-grained knowledge retrieval. As shown in Figure~\ref{fig:hyper}(c) and \ref{fig:hyper}(f), performance remains consistently high. This stability confirms that the angular violation score effectively prioritizes semantically relevant fine-grained neighbors, making the ranking robust to an enlarged retrieval window.

\subsubsection{Hyper-parameters in Dual-path Fusion}
Figure~\ref{fig:hotmap} illustrates the classification performance as we vary the fusion weights $\beta$ and $\alpha$ within the range of $\{0.1, 0.3, 0.5, 0.7, 0.9, 1.0\}$. The heatmap results indicate that feature-level fusion (controlled by $\beta$) serves as the primary factor. In particular, WikiCS benefits from substantial knowledge injection, corresponding to larger values of $\beta$, whereas CiteSeer favors a more conservative degree of knowledge integration. In contrast, structure-level fusion (controlled by $\alpha$) consistently improves performance across a wide range of values. The presence of broad high-performance regions further demonstrates the robustness of HyRAG with respect to hyper-parameter variations.

\section{Conclusion}
By analyzing the geometric properties of Euclidean space, we identify two fundamental limitations of Euclidean-based RAG frameworks: the loss of semantic granularity and the emergence of hubness.
To overcome these limitations, we propose HyRAG, a hyperbolic retrieval-augmented generation framework that enhances the generalization capability of GFMs. By integrating hyperbolic knowledge indexing, multi-granularity retrieval, and dual-path fusion, HyRAG enables geometry-aware utilization of global semantic anchors and local semantic nuances. Extensive experiments on multiple graph benchmarks demonstrate that HyRAG consistently improves zero-shot inference performance without any task-specific fine-tuning, suggesting hyperbolic RAG as a promising direction for building robust and generalizable graph foundation models.

\section*{Acknowledgments}
This work is supported by National Natural Science Foundation of China No. 62406313, Postdoctoral Fellowship Program of China Postdoctoral Science Foundation, Grant No. YJB20250283.


\bibliographystyle{ACM-Reference-Format}
\balance
\bibliography{sample-base}

\appendix

\section{Datasets}\label{sec:datasets}
In this paper, we employ seven widely used graph datasets, as summarized in Table~\ref{tab:dataset}. A detailed introduction to these datasets is provided below.

\begin{table}[htp]
    \centering
    \caption{Statistics of the datasets.}
    \label{tab:dataset}
    \begin{tabular}{c|ccccc}
    \toprule
   Dataset &  \#Nodes&\#Edges   & Domain   &  \#Category  \\ \midrule
    Cora& 2,708 & 5,429 &Academic & 7 \\
    CiteSeer & 3,186 & 4,277    &Academic  & 6 \\
    Ele-Photo  & 48,362  & 500,928  &E-commerce& 12 \\
    Ele-Computers  & 87,229  & 721,081  &E-commerce& 10 \\
    Books-History& 41,551  & 358,574  &E-commerce& 12 \\
    WikiCS     & 11,701  & 215,863  &Wikipedia & 10      \\
    Instagram  & 11,339  & 144,010  &Social    & 2        \\
        \bottomrule
    \end{tabular}
\end{table}

\textbf{Cora}~\cite{cora} is a citation network consisting of 2,708 nodes and 5,429 edges, where each node represents a scientific publication and each edge indicates a citation between publications. The textual attribute of each node includes the title and abstract of the corresponding publication. Node categories span seven research areas: case-based reasoning, genetic algorithms, neural networks, probabilistic methods, reinforcement learning, rule learning, and theory.

\textbf{CiteSeer}~\cite{citeseer} is also a citation network, consisting of 3,186 nodes and 4,277 edges. Each node represents a scientific publication, categorized into one of six research areas: Agents, Machine Learning, Information Retrieval, Database, Human-Computer Interaction, and Artificial Intelligence.

\textbf{Ele-Photo}~\cite{cstag} is derived from the Amazon Electronics dataset, where each node represents an electronic product and each edge denotes frequent co-purchases or co-views. The text attribute of each node comes from the most upvoted user review or a random review if no such votes are available. 
The classification task involves assigning each product to one of 12 categories.

\textbf{Ele-Computers}~\cite{cstag}, also from Amazon Electronics, shares the same structure but classifies products into 10 categories. Text attributes follow the same rule as Ele-Photo.

\textbf{Books-History}~\cite{cstag} extracted from the Amazon-Books dataset, focuses on history books. Nodes represent books with edges indicating co-purchases or co-views. Text attributes include titles and descriptions. The classification involves 12 categories.

\textbf{WikiCS}~\cite{wikics} dataset is a Wikipedia-based dataset designed for benchmarking Graph Neural Networks. It comprises 10 classes, each corresponding to a branch of computer science.

\textbf{Instagram}~\cite{instagram} is a social network dataset where nodes represent users and edges indicate the following relationships. The prediction task is to classify each user as either commercial or normal.

\section{Implementation Details} \label{sec:imp}
We adopt the pre-trained GraphCLIP model as the backbone, which has been trained on the ogbn-ArXiv, ArXiv\_2023, PubMed, ogbn-Products, and Reddit datasets. All experiments are conducted with a batch size of 32 on one NVIDIA V100 GPU. 
In the HKI module, we employ all-MiniLM-L6-v2 \cite{minilm} as the language model, and the hyper-parameters are configured as follows: $\epsilon = 0.2$, $\lambda_a = 0.5$, and $\lambda_\text{reg} = 0.1$. 
For the MR module, we utilize LLaMA-3.1-8B-Instruct \cite{llama} as the LLM to generate the corresponding coarse-grained and fine-grained queries. The number of retrieved central entities
 $k$ is selected from $\{1, 3, 5, 7, 9\}$, the diversity factor $\gamma$ is tuned within $\{0.1, 0.3, 0.5, 0.7, 0.9, 1\}$, and the number of retrieved neighborhood entities in fine-grained knowledge retrieval $k'$ is chosen from $\{5, 10, 15, 20\}$. Finally, in the DF module, the temperature coefficient $\tau$ is set to 5, while the feature-level fusion hyperparameter $\beta$ and the structure-level fusion hyperparameter $\alpha$ are both tuned within $\{0.1, 0.3, 0.5, 0.7, 0.9, 1\}$.
 We report the mean accuracy and standard deviation over three runs with different random seeds.

\end{document}